\documentclass[aps,prl,twocolumn,floatfix,twocolumn,citeautoscript,showpacs,notitlepage,superscriptaddress]{revtex4-1}

\usepackage[english]{babel}
\usepackage[pdftex,colorlinks=false]{hyperref}

\addto\captionsenglish{}
\usepackage{latexsym}
\usepackage{amsfonts}
\usepackage{amsmath,amssymb,amsthm}
\usepackage{bbm}
\usepackage{mathtools}
\usepackage[dvipsnames]{xcolor}
\usepackage{graphicx}
\usepackage{bm}
\usepackage{multirow}
\usepackage{tikz}
\usetikzlibrary{shapes}
\usepackage{txfonts}
\usepackage{csquotes}
\usepackage[capitalize]{cleveref}
\usepackage{lipsum}
\usepackage{blindtext}
\usepackage{subfiles}
\usepackage{amssymb}
\usepackage{amsmath}
\usepackage{eurosym}

\let\origunderline\underline
\renewcommand{\underline}[1]{\origunderline{\smash{#1}}}

\newcommand{\bs}[1]{{\boldsymbol{#1}}}

\renewcommand{\i}{\mathrm{i}}

\renewcommand{\mathbb}{\mathbbm}

\renewcommand{\dag}{\dagger}
\newcommand{\nodag}{{\vphantom\dag}}

\definecolor{ex_fill}{HTML}{b7da27}
\definecolor{ex_border}{HTML}{8a983f}

\newrobustcmd*{\greenSquare}{\tikz[baseline=0.13ex]{
    \filldraw[draw=ex_border,fill=ex_fill, thick] (0,0) rectangle (0.2cm,0.2cm);}}

\newrobustcmd*{\greenHexagon}{\tikz[baseline=-0.53ex]{
    \node[draw=ex_border, fill=ex_fill,regular polygon, regular polygon sides=6,
              inner sep=0.07cm, thick] at (0,0){};}}

\newrobustcmd*{\greenCircle}{\tikz[baseline=-0.53ex]{
    \filldraw[draw=ex_border,fill=ex_fill, thick] (0,0) circle (0.1cm);}}
\newrobustcmd*{\greenTriangleUp}{\tikz[baseline=0.13ex]{\filldraw[
    draw=ex_border,fill=ex_fill, thick] (0,0) -- (0.2cm,0) -- (0.1cm,0.2cm) --
    cycle;}}
\newrobustcmd*{\greenTriangleDown}{\tikz[baseline=-1.23ex]{\filldraw[
    draw=ex_border,fill=ex_fill, thick] (0,0) -- (0.2cm,0) -- (0.1cm,-0.2cm) --
    cycle;}}

\definecolor{dr}{rgb}{0, 0, 1}

\definecolor{sr}{rgb}{0.6, 0.2, 0.7}

\definecolor{mk}{rgb}{1, 0, 0}

\definecolor{rt}{rgb}{0, 1, 0}

\definecolor{js}{rgb}{0, 0.5, 0.5}

\definecolor{pw}{rgb}{0.5, 0.5, 0}

\begin{document}


    \title{Incommensurate magnetic order: A fingerprint for electronic correlations in hole-doped cuprates}

    \author{Michael~Klett}
    \email{michael.klett@uni-wuerzburg.de}
    \affiliation{Institute for Theoretical Physics, University of W\"urzburg,
        D-97074 W\"urzburg, Germany}
    \affiliation{Würzburg-Dresden Cluster of Excellence ct.qmat, University of W\"urzburg,
        D-97074 W\"urzburg, Germany}
    
    \author{Jacob~Beyer}
    \affiliation{Institute for Theoretical Physics, University of W\"urzburg,
        D-97074 W\"urzburg, Germany}
    \affiliation{Institute for Theoretical Solid State Physics, RWTH Aachen
        University, D-52062 Aachen, Germany}
    \affiliation{School of Physics, University of Melbourne, Parkville,
        VIC 3010, Australia}

    \author{David~Riegler}   
    \email{david.riegler@kit.edu}
    \affiliation{Institute for Theoretical Physics, University of W\"urzburg,
        D-97074 W\"urzburg, Germany}
	\affiliation{Institute for Theory of Condensed Matter, Karlsruhe Institute of Technology, D-76128 Karlsruhe, Germany}

    \author{Jannis~Seufert}
    \affiliation{Institute for Theoretical Physics, University of W\"urzburg,
        D-97074 W\"urzburg, Germany}
     \affiliation{Würzburg-Dresden Cluster of Excellence ct.qmat, University of W\"urzburg,
        D-97074 W\"urzburg, Germany}
    
	\author{Peter~W\"olfle}
	\affiliation{Institute for Theory of Condensed Matter, Karlsruhe Institute of Technology, D-76128 Karlsruhe, Germany}
    \affiliation{Institute for Quantum Materials and Technologies, Karlsruhe Institute of Technology, D-76021 Karlsruhe, Germany}

    \author{Stephan~Rachel}
    \affiliation{Würzburg-Dresden Cluster of Excellence ct.qmat, University of W\"urzburg,
        D-97074 W\"urzburg, Germany}
    \affiliation{School of Physics, University of Melbourne, Parkville,
        VIC 3010, Australia}

    \author{Ronny Thomale}
    \email{rthomale@physik.uni-wuerzburg.de}
    \affiliation{Institute for Theoretical Physics, University of W\"urzburg,
        D-97074 W\"urzburg, Germany}
    \affiliation{Würzburg-Dresden Cluster of Excellence ct.qmat, University of W\"urzburg,
        D-97074 W\"urzburg, Germany}

    \date{\today }

    \begin{abstract}
    Intertwined charge and magnetic fluctuations in high-$T_\text{c}$ copper oxide
    superconductors (cuprates) are hypothesized
    to be a consequence of their correlated electronic
    nature. Among other observables, this is apparent in the
    doping dependence of
    incommensurate magnetic order, known as the Yamada
    relation (YR). We analyze the Hubbard model to challenge the
    universality of YR as a function of interaction strength $U$ through
    Kotliar-Ruckenstein slave-boson (SB) mean-field theory and truncated
    unity functional renormalization group (TUFRG). While TUFRG tends to lock in
    to a doping dependence of the incommensurate magnetic ordering
    vector obtained for the perturbative weak-coupling limit, SB
    not only exhibits an enhanced sensitivity upon a variation of $U$ from weak to
    strong coupling, but also shows good agreement with experimental
    data. It supports the placement of weakly hole-doped
    cuprates in the intermediate-to-strong coupling regime. 
    \end{abstract}

\maketitle

    \textit{Introduction} --- Due to its condensed
    simplification and yet remarkable complexity in terms of quantum
    phases of matter it gives rise to, the Hubbard model is considered
    the drosophila of correlated condensed matter physics. With respect to copper oxide superconductors
    (cuprates), the Hubbard model suggests itself to even capture many
    rather subtle features of magnetic/charge order and
    superconductivity. Despite its elementary formulation and
    robust microscopic foundation, the Hubbard model has proven
    immensely challenging to solve, rendering its solution in two and
    more spatial dimensions a cornerstone in the quest for understanding
    and predicting the properties of cuprate-type materials. For
    instance, recent strives for finding the exact ground state are just one of many ongoing concerted efforts 
    by the scientific community to determine the properties of the
    Hubbard model
    \cite{Schaefer_2021}.

    One of the many intriguing phenomena observed in the Hubbard model is the emergence
    of incommensurate magnetic spiral order, in particular for hole
    doping away from pristine half filling.
    Such a spiral is characterized by a periodic modulation of the
    magnetic moments, where the modulation does not have an integer
    number of wavelengths within the underlying lattice. While such
    phenomena are readily interpreted from a weak coupling itinerant
    limit where the nesting, and hence the preferred magnetic ordering,
    continuously detunes a function of hole-doped Fermiology, the
    microscopic attribution of doping-dependent incommensurate magnetic order is
    much more difficult in the intermediate to strong coupling
    regime. Approaching such a parametric domain of the
    Hubbard model from a holistic perspective calls for methods, which are
    (i) not limited by real space
    cluster size in their resolution of incommensurability, (ii)
    capable of including the dynamical interplay of charge and spin
    fluctuations, and (iii) feature a comprehensive tunability of
    $U$. The truncated unity functional renomalization group
    (TUFRG)
    \cite{beyer2022a,lichtenstein2017,metzner2012}
    suggests itself as such a candidate method. Rooted in an RG
    scheme of electronic diagrammatic resummation, it is
    consistently formulated  in momentum space amenable to
    incommensurability phenomena and couples charge and spin
    fluctuations at variable interaction vertex strength
    $U$. Furthermore, the Kotliar-Ruckenstein slave-boson (SB) method
    \cite{kotliar_new_1986}, when
    generalized to its spin-rotation invariant form
    \cite{woelfle_spin_rotation_1989}, provides a highly suitable mean
    field method to track incommensurate magnetic order at arbitrary
    doping and interaction strength $U$. In particular, as shown recently in a methodological revival and makeup of the method, SB can provide a detailed fluctuation profile around any kind of reference mean field~\cite{Hubbard_Wuerzburg}, resolve the topology of strong correlated electron systems~\cite{PhysRevB.101.161112}, and reach a remarkably accurate
    modelling of experimental evidence on incommensurate charge order in electron-doped cuprates~\cite{Riegler_2023}.



    In this paper we do two things: First, we employ TUFRG and SB to
    analyze the hole-doping dependence of magnetic order in the Hubbard
    model. As TUFRG approaches the intermediately coupled electronic
    regime from the perturbative weak-coupling limit, while SB additionally features
    certain controlled asymptotic limits in the complementary strongly-coupled limit,
    we assume our combined assessment from both methods to accurately capture a
    broad range of interaction strengths. Moreover, both methods allow
    extracting similar observables, and hence enable a comparison on equal
    footing. Second, we challenge our analysis of incommensurate
    magnetic order with experimental evidence. In particular,
    this concerns the Yamada relation in hole-doped La-based cuprate
    materials \cite{Yamada_1998}, which correlates
        incommensurate magnetic spiral vectors with the doping
        level~\cite{comin2016resonant,Lee_2022}.
    We find that the intermediate-to-strong coupling regime, as resolved
    through SB, provides the best accordance with experimental data by a
    significant margin.
    

    \begin{figure*}[t!]
        \includegraphics[width=1\textwidth]{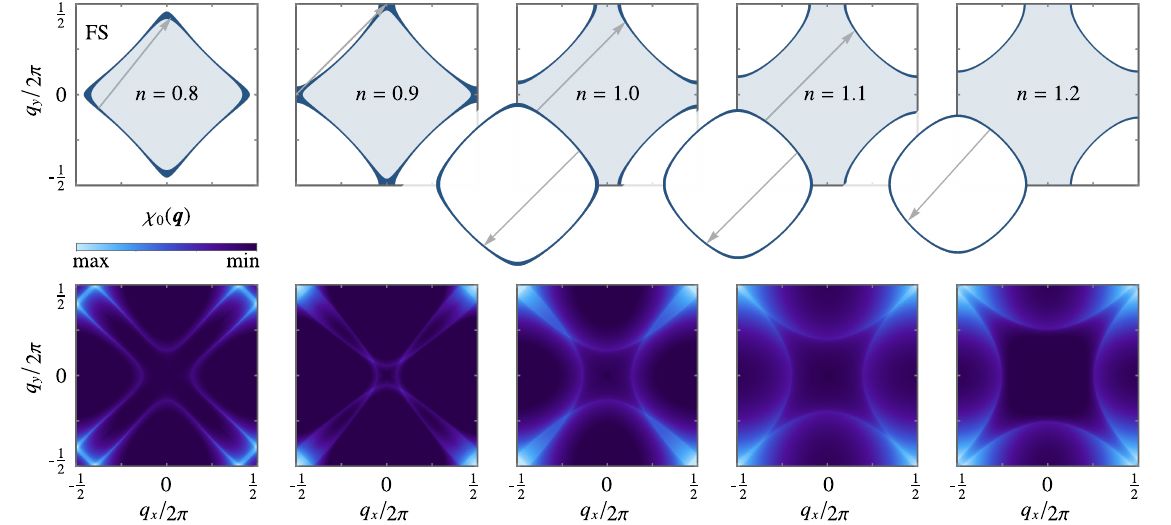}
        \caption{Upper row: FS of the single band Hubbard
        model \cref{eq:Hamiltonian} in the zero interaction limit at different
        fillings. The width of the FS indicates the density of states, the
        arrow indicates the maximally nested vector of the bare
        susceptibilities.
        Lower row: Corresponding bare susceptibilities for the above FS.
        We can see that for sufficient doping the transfer momentum with
        maximal support becomes incommensurate, \textit{e.g.}, at $n=0.8$ the nesting wave vector $\boldsymbol{q}/2\pi$ is shifted away from $\big(\tfrac{1}{2},\tfrac{1}{2}\big)$.
        \label{fig:FS}
        }
    \end{figure*}

    \newpage
    \textit{Model} ---
    We analyze the one-band $t$\,--\,$t'$\,--\,$U$\ Hubbard
    model on the two-dimensional square lattice defined by
    \begingroup
    \allowdisplaybreaks
        \begin{align}\label{eq:Hamiltonian}
        \begin{split}
        H= &- \sum_{\sigma=\uparrow,\downarrow}
        \left(
        t^\nodag_{\vphantom i} \sum_{\langle i,j \rangle_1}
                \, c_{i,\sigma }^{\dagger}c^\nodag_{j,\sigma}
        + t{\vphantom i}'  \sum_{\langle i,j \rangle_2}
                c_{i,\sigma }^{\dagger}c^\nodag_{j,\sigma}
        +\mathrm{h.c.}
        \right) \\
        &- \sum_{\sigma=\uparrow,\downarrow} \mu_0 \sum_i  c_{i,\sigma }^\dagger c_{i,\sigma}^\nodag
        +U\sum_{i} c_{i,\uparrow }^\dagger c_{i,\uparrow}^\nodag
            c_{i,\downarrow }^\dagger c_{i,\downarrow }^\nodag 
        \, , 
        \end{split}
        \end{align}
    \endgroup
    where the operator $c_{i,\sigma}^{\dagger }$ creates an electron
    with spin $\sigma \in \{\uparrow,\downarrow\} $ at site $i$.
    Moreover, $\langle i,j \rangle_n$ denotes a $n$\textsuperscript{th} nearest
    neighbor pair, and $\mu_0$ is the chemical potential.
    To introduce particle-hole asymmetry as experimentally observed
    we employ $t'/t=-0.15$ throughout the paper,
    which is a generic parameter in cuprate systems \cite{doi:10.1080/00018732.2014.940227,Hirayama_2018,PhysRevB.72.054519,honerkamp2001}
    and measure energy in units of $t$.    

    \textit{Bare Susceptibility} ---
    To give an initial intuition of the model and its propensity towards incommensurate order, we investigate the bare susceptibility $\chi_0$ which -- in the absence of orbital dependencies -- is equivalent to the random-phase approximation (RPA) spin susceptibility $\chi_s$ \cite{Hochkeppel_2008,Altmeyer_2016, Dürrnagel2022}
    \begin{equation}
      \chi_0(\bs q) = \sum_{\bs k} \frac{n_\text{F}(\epsilon_{\bs k}) - n_\text{F}(\epsilon_{\bs k + \bs q})}{\epsilon_{\bs k + \bs q} - \epsilon_{\bs k} + \i 0^+}, \quad \chi_s(\bs q) = \frac{\chi_0(\bs q)}{1 - U \chi_0(\bs q)},
    \end{equation}
    with $n_\text{F}$ representing the Fermi-Dirac distribution and $\epsilon_{\bs k}$ the energy eigenvalues of Eq.~\eqref{eq:Hamiltonian} with $U \to 0$.
    The results reveal nesting properties
    of the undressed electrons, which stem from the Fermiology
    of the underlying band structure.
    The leading nesting vectors for a range of fillings are shown in the upper row of \cref{fig:FS} as arrows,
    while the lower row depicts the corresponding susceptibilities $\chi_0(\bs q)$.
    Already here, we observe the broken electron-hole symmetry:
    while the maxima of the susceptibility remain close to
    $\bs q / 2\pi =\big(\tfrac{1}{2},\tfrac{1}{2}\big)$ for electron doping, they shift
    significantly on the
    hole-doped side. 
    Unsurprisingly, for large dopings the warping of the Fermi surface (FS) induces incomensurabilities
    in the correlations, which we will compare to the TUFRG and SBMF results. 

    \textit{Truncated-Unity FRG} ---
    To investigate weak-to-intermediate coupling strengths, we employ the TUFRG.
    This method emphasizes the diagrammatic transfer momenta of the three channels:
    particle-hole, crossed particle-hole, and particle-particle, and thus captures
    occurring instabilities of all three.
    The secondary -- slowly varying -- momentum dependencies are projected onto
    basis functions $\varphi = e^{\i \bs k \bs r}$,
    with $\bs r$ lattice sites, to reduce computational cost.
    The resulting TUFRG flow equations are solved by introducing an artificial
    regulator, interpolating between the high-scale no-fluctuation regime and
    the full-fluctuation regime at low scale.
    Reformulating the problem as derivatives w.r.t.~this artificial scale $\Lambda$, we
    obtain an infinite hierarchy of coupled ordinary differential equations.
    Focusing on the two-particle terms, we can integrate the equations to obtain the
    effective two-particle interaction at the phase transition
    \cite{beyer2022, metzner2012, platt2013}.
    Neglecting further terms implies that the FS 
    does not experience any additional warping or renormalization.

    \begin{figure*}[t!]
        \includegraphics[width=1\textwidth]{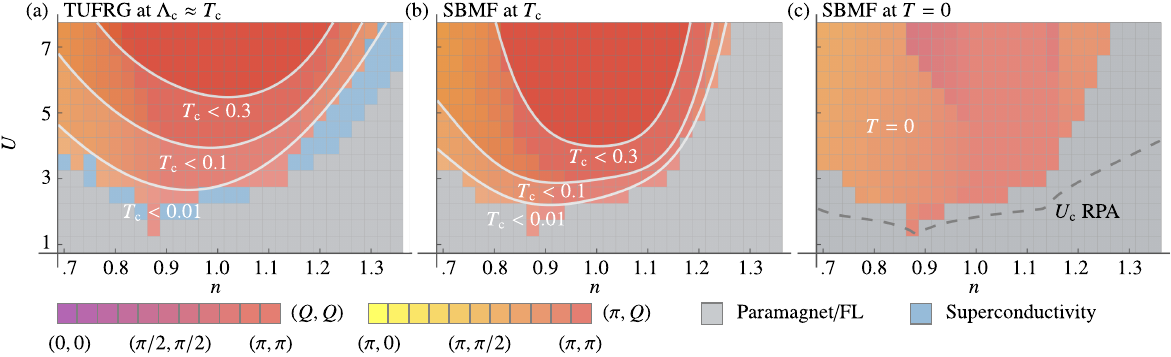}
        \caption{
            Phase diagrams from TUFRG at critical temperature $T_\text{c}$ (a),
            SBMF at critical temperature $T_\text{c}$ (b),
            and SBMF at zero temperature $T=0$ (c).
            The color gives the phase prediction offered by the respective
            methods, note that only TUFRG is capable of observing superconducting
            pairing instabilities.
            The shade of red denotes the degree of incommensurability, with
            purple measured along the $\Gamma$-$\mathrm{M}$ line and yellow measured
            along the $\mathrm{X}$-$\mathrm{M}$ line.
            Highlighted by white contour lines in the filling interaction
            parameter space are fits for isothermal lines of the transition
            temperature $T_\text{c}$.}
        \label{fig:PS}
    \end{figure*}

    The initial condition is given by the bare action of
    the system at high temperature, while termination is indicated
    by a divergence of vertex elements and thereby defines the critical scale of $\Lambda_{\text c} \approx T_{\text c}$.
    The nature of the phase transition implied by the divergence, we ascertain by means    of a simple eigendecomposition in the formfactor space $\varphi$ of the
    effective two-particle interaction
    $V_{\varphi\varphi'}^{\Lambda_{\text c}}(\bs Q)$.
    The subspace of the leading eigenvalues determines the ordering vector $\bs Q$.
    If we do not encounter a divergence above a threshold scale,
    we determine the system to be a Fermi liquid (FL).

    \textit{Slave-boson mean-field approximation} --- 
    To complement the weak-to-intermediate coupling captured by the TUFRG, 
    we apply the spin
    rotation invariant Kotliar-Ruckenstein slave-boson representation ~\cite{kotliar_new_1986,woelfle_spin_1992}. This approach non-perturbatively addresses local and short-ranged density-density interactions by transforming fermionic operators $c$ into new representations involving bosonic auxiliary fields $\psi$ and pseudofermions $f$:
        \begin{align}
        c_{\alpha} = {Z}^{\phantom\dagger}_{\alpha \beta}[\psi] \, f^{\phantom\dagger}_{\beta} \mathcal P.
        \end{align}
    Here, $\alpha$ and $\beta$ are sets of quantum numbers uniquely labeling local states. The transformation between physical electrons and pseudofermions is governed by a renormalization matrix $Z$, with bosonic operators as entries \cite{woelfle_spin_1992} chosen such that the results of the Gutzwiller variational method \cite{Gutzwiller_1963} are recovered on the mean field level. Additionally we have to impose constraints to restore the original Hilbert space by means of projectors $\mathcal P$. This results in quadratic auxiliary field terms for local quartic interactions and a natural kinetic energy scale renormalization. Pseudofermions represent quasiparticles, yielding a renormalized Fermi liquid at the mean-field level, while the condensed bosonic fields can be interpreted as probability amplitudes in the local configuration space.


    To determine the ground state, we employ a static mean-field (MF) ansatz. Charge degrees of freedom are approximated as spatially uniform $\langle n_i \rangle = n$, while spin degrees are constrained to a spiral $\langle \bs S_i \rangle = S[\psi]\left( \cos(\bs Q \, \bs r_i), \sin(\bs Q \, \bs r_i),0 \right)$ with ordering vector $\bs Q$, where $ S \neq 0$ indicates magnetic order. Utilizing block diagonality, the MF Hamiltonian is represented by a $2\times 2$ matrix for arbitrary $\bs Q$, facilitating the comparatively easy exploration of incommensurabilities \cite{Fresard_1991,Hubbard_Wuerzburg,Seufert_2021,Riegler_2023}. The MF ground state is determined by the saddle point of the Free Energy, with the constraints enforced as thermodynamic averages.


    \begin{figure*}[t!]
        \includegraphics[width=1\textwidth]{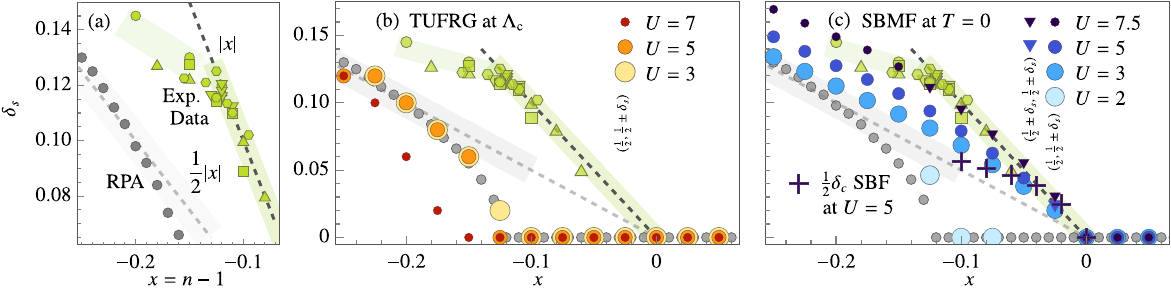}
        \caption{(a) We compare the RPA susceptibility (gray) to experiments (green),
        as well as the (b) TUFRG (red) and (c) SB (blue) calculations respectively.
            The experiments exhibit the Yamada relation ($\delta_s = |1-n|$)
            in the hole doped regime, reproduced by SBMF for increasing
            interaction, including a decreasing slope around
            $\delta_s\approx0.1$.
            In contrast, TUFRG recovers the weak coupling line predicted by the
            RPA susceptibility, but fails
            at high interaction strengths $U > 2$.
            The experimental RXS \cite{comin2016resonant} (charge channel, using $2\delta_s = \delta_c$) and neutron-scattering \cite{Yamada_1998} (spin channel) measurements data is taken from
            Ref.~\cite{Fink_2009} La$_{1.8-x}$Eu$_{0.2}$Sr$_x$CuO$_4$(\greenSquare),
            Ref.~\cite{Yamada_1998} La$_{2-x}$Sr$_{x}$CuO$_{4}$ (\greenTriangleUp),
            Ref.~\cite{Wu_2019,Croft_2014,Christensen_2014} La$_{2-x}$Sr$_x$CuO$_4$ (\greenTriangleDown),
            Ref.~\cite{Tranquada_1997} La$_{1.6-x}$Nd$_{0.4}$Sr$_x$Cu$_2$O$_{4}$ (\greenCircle),
            Ref.~\cite{Abbamonte_2005,Huecker_2013} La$_{2-x}$Ba$_x$CuO$_4$(\greenHexagon).
            \label{fig:QPlot}
        }
    \end{figure*}

    \textit{Comparison of phase diagrams} ---
    Fig. \ref{fig:PS} illustrates the $U$-filling phase diagrams obtained by the TUFRG (a) and SBMF (b) methods at the critical temperature $T_\text{c}$ of the magnetic or superconducting transition. Notably, a qualitative similarity exists between the two methods, particularly where superconductivity is suppressed by interactions, validating TUFRG's continued applicability. SBMF, being agnostic to superconducting phases, exclusively identifies magnetic phase transitions.
    We mention in passing that although thermal fluctuations are known to destroy long range order in strictly two-dimensional systems (Mermin-Wagner theorem), the growth of the correlation length may trigger a phase transition in the (experimentally studied) three-dimensional systems, even for very small inter-layer coupling. We identify the transition found in SBMF with such a transition.

    Consistent with prior large-$T$ SB studies of the Hubbard model \cite{Hubbard_Wuerzburg}, we observe incommensurate spiral magnetism for hole-doping $\bs Q/2\pi=\big(\frac12,\frac12\pm\delta_s\big)$ and a commensurate antiferromagnetic (AFM) domain for electron-doping around half-filling, which matches the TUFRG results.   
    The equivalence in the characteristics of phase diagrams validates our calculations for the weak-to-intermediate coupled regime, \textit{i.e.}, $U \lesssim W/2$. For larger interactions, deviations in isothermal lines of the critical temperature emerge. TUFRG, due to truncation, generally underestimates critical scales for strong interactions and is in its viability is constrained to $U \approx W/2$, while SBMF, lacking fluctuation corrections, predicts higher transition temperatures \cite{Sabiryanov_1997,Peng_2008}.     

    Additionally, a SBMF phase diagram (c) at zero temperature $T=0$ reveals incommensurate spiral magnetism with wavevecors  $\bs Q/2\pi=\big(\frac12,\frac12\pm\delta_s\big)$ and $\bs Q/2\pi=\big(\frac12\pm\delta_s,\frac12\pm\delta_s\big)$ as well as a more pronounced incommensurability parameter $\delta_s$. Notably, SBMF studies at low $T$ suggest phase separation of commensurate and incommensurate antiferromagnetism around half-filing \cite{igoshev2013,igoshev2015,Seufert_2021}.
    However, recent methodological strives have shown, that long-ranged density-density interactions significantly reduce the phase-separated domain \cite{Riegler_2023}. Applying ab initio values for the interactions \cite{Hirayama_2018}, we detect phase separation for $x\lesssim 0.05$ and a continuous evolution of the incommensurability parameter $\delta_s$ as a function of doping for $x \gtrsim 0.05$ in line with the Yamada behavior.   

    We contrast our results with the critical interaction $U_\text{c}$ for the spin channel predicted by RPA, which qualitativly follows the results found using TUFRG and the corresponding incommensurability parameters can be inferred from \cref{fig:QPlot}. Since RPA only involves the particle-hole ladder-type diagrams and does not account for all scattering processes, the overall critical scale of the system is underestimated, especially for dopings far away from the van Hove singularity at $n \approx 0.87$, where the large DOS at the Fermi level dominates the emerging phase transition.

    To connect to experimental results, we remember that the Hubbard model is often used as a
    simple model of cuprate systems.
    This relies on the electron's confinement to the copper-atoms for the electron-doped regime
    and the Zhang-Rice-singlet \cite{zhang1988} for the hole-doped regime.
    Despite its inherent simplifications and approximations, Eq.~\eqref{eq:Hamiltonian} has proven
    
    to be a valuable theoretical model for gaining insights
    into the behavior of high-temperature superconductors.
    To contextualize the above presented results we compare them to spin
    incommensurablities measured in the La-based cuprate family. 

    \textit{Experimental results and Yamada relation} --- Incommensurability is not merely a theoretical concept; it manifests in high-$T_\text{c}$ cuprates. Yamada \textit{et al.} \cite{Yamada_1998} established this through neutron-scattering experiments on La$_{2-x}$Sr$_{x}$CuO$_4$ and showed that the low-energy spin-fluctuation peak shifts from $\big(\frac12, \frac12\big)$ to $\big(\frac12 \pm \delta_s, \frac12\big)$ and $\big(\frac12, \frac12 \pm \delta_s\big)$ beyond a critical hole doping ($|x| \gtrsim 0.05$). Further studies additionally revealed an incommensurate spin-density wave, along $\big(\frac12 \pm \delta_s, \frac12 \pm \delta_s\big)$, which rotates upon increased doping by $45^\circ$ back to the order type reported by Yamada \textit{et al.}, resembling the findings of the SBMF phase diagram at constant temperature~\cite{Comin_2014,PhysRevB.97.155144}. The incommensurability parameter $\delta_s$ approximately follows the relation $\delta_s =|x|$, known as the Yamada relation.
    
    Further investigations on related materials \cite{hucker2011,Croft_2014} show coexistence of charge density wave and incommensurate magnetic order, where the wave vectors exhibit a simple relationship $\delta_c = 2\delta_s$. The Yamada relation is therefore sometimes expressed as in terms of the charge incommensurability $\delta_c =2|x|$. The interplay of incommensurate charge and spin order in the Hubbard model has been studied on finite size systems using the above SB method with similar results for $\delta_s$\cite{raczkowski2006interplay}. A recent discussion of the quality of the above SB method can be found in \cite{Philoxene_2022} (part II.C.1.). We explore charge instabilities atop commensurate magnetic order by analyzing the charge susceptibility within the SB representation and the fluctuation (SBF) formalism from Refs.~\cite{Seufert_2021,Riegler_2023}. Our results show no charge instabilities within the commensurate AFM domain but indicate a charge incommensurability accompanying the spin incommensurability, roughly following the previously mentioned behavior, compare Fig. \ref{fig:QPlot}(c). In contrast to electron-doping \cite{Seufert_2021,Riegler_2023}, we find $\delta_c$ directly tied to $\delta_s$. However, as the current formalism lacks the ability to calculate fluctuations around incommensurate ground states, these results are a reliable approximation only for small dopings and intermediate values of interaction.
        
    Note, that other cuprate families, \textit{e.g.}, Y- and Bi-based compounds, do not follow the Yamada relation and realize other types of spin and charge incomensurablilties \cite{comin2016resonant}. For comparison experimental measurements \cite{Tranquada_1997,Yamada_1998,Abbamonte_2005,Fink_2009,Huecker_2013,comin2016resonant,Croft_2014,Christensen_2014,Wu_2019,Lee_2022} of the Yamada dependence $\delta_s
    \approx |x|$ stemming for charge and spin measurements are included in \cref{fig:QPlot}.

    \textit{Incommensurability as a function of filling} --- In \cref{fig:QPlot}(a), we illustrate the disparity between the observed Yamada line and the bare susceptibility calculation. The reduced incommensurability in the bare calculations can be attributed to the absence of FS warping effects, \textit{i.e.}, self-energy contributions. \cref{fig:QPlot}(b) showcases the alignment of TUFRG calculations with the RPA result, especially up to couplings of $U \approx W/2$, beyond which TUFRG is deemed somewhat unreliable \cite{Hille_2020}. This underscores the presumed connection between FS behavior, as represented by the bare susceptibility, and TUFRG calculations.
    In contrast, SBMF results not only replicate the Yamada line at high interaction values ($U \gtrsim W/2$) but also capture the low coupling behavior seen in the bare susceptibility. Noteworthy is the presentation of these calculations at a constant temperature, resembling an experimental setup more closely. At the critical temperature, SBMF does not coincide with the Yamada line. We deduce from these findings that the renormalization of the FS, the emergence of the magnetic phase beyond the paramagnetic-magnetic transition, and substantial bare couplings are all essential factors for the establishment of the Yamada regime.


    \textit{Conclusion} --- We analyze the one-band Hubbard model to explore the dependence of
    incommensurate magnetic order on doping and interaction strength
    $U$ within a wide parameter space.
    Comparing the results of our chosen SB and TUFRG approaches,
    both methods reproduce the Fermiology-driven magnetic instability
    already visible from the bare susceptiblity. At
    intermediate-to-large $U$, it is SB that most accurately
    reproduces a magnetic footprint akin to the Yamada relation for
    hole-doped cuprates. As this is a comprehensive
    theoretical reproduction of this experimental phenomenology, one
    might expect a renaissance of SB methodology to address
    intricate many-body phenomena in strongly
    correlated electron systems.
   
    \textit{Acknowledgments}---
    The authors thank Carsten Honerkamp, Matteo Dürrnagel, Hendrik Hohmann, Jonas Issing, Tobias M\"uller, and
    Tilman Schwemmer for helpful
    discussions.
    The work in W\"urzburg is funded by the Deutsche Forschungsgemeinschaft
    (DFG, German Research Foundation) through Project-ID 258499086 - SFB 1170
    and through the W\"urzburg-Dresden Cluster of Excellence on Complexity and
    Topology in Quantum Matter -- \textit{ct.qmat} Project-ID 390858490 - EXC
    2147.
    J.B. is funded by the DFG through RTG 1995.
    We acknowledge HPC resources provided by the Erlangen National High
    Perfomance Computing Center (NHR@FAU) of the
    Friedrich-Alexander-Universität Erlangen-Nürnberg (FAU).
    NHR funding is provided by federal and Bavarian state authorities. NHR@FAU
    hardware is partially funded by the DFG – 440719683. P.W. acknowledges support through a Distinguished Senior Fellowship of Karlsruhe Institute of Technology. S.R. acknowledges support from the Australian Research Council through Grant No.\ FT180100211.


    \bibliographystyle{apsrev4-2}
    \bibliography{Main}
\end{document}